\documentclass[fleqn,12pt,twoside]{article}
\usepackage[headings]{espcrc1}

\readRCS
$Id: espcrc1.tex,v 1.2 2004/02/24 11:22:11 spepping Exp $
\usepackage{graphicx}
\usepackage[figuresright]{rotating}

\newcommand{\AmS}{{\protect\the\textfont2
  A\kern-.1667em\lower.5ex\hbox{M}\kern-.125emS}}

\title{First measurement of the $\rho$ spectral function in nuclear collisions}
\author{S. Damjanovic for the NA60 Collaboration:\\
R.~Arnaldi$^{10}$, 
R.~Averbeck$^{9}$, 
K.~Banicz$^{2,4}$, 
J.~Castor$^{3}$, 
B.~Chaurand$^{7}$, 
C.~Cical\`o$^{1}$, 
A.~Colla$^{10}$, 
P.~Cortese$^{10}$, 
S.~Damjanovic$^{4}$, 
A.~David$^{2,5}$, 
A.~De~Falco$^{1}$, 
A.~Devaux$^{3}$, 
A. Drees$^{9}$, 
L.~Ducroux$^{6}$, 
H.~En'yo$^{8}$, 
A.~Ferretti$^{10}$, 
M.~Floris$^{1}$, 
P.~Force$^{3}$, 
N.~Guettet$^{2,3}$, 
A.~Guichard$^{6}$, 
H.~Gulkanian$^{11}$, 
J.~Heuser$^{8}$, 
M.~Keil$^{2,5}$, 
L.~Kluberg$^{2,7}$, 
C.~Louren\c{c}o$^{2}$, 
J.~Lozano$^{5}$, 
F.~Manso$^{3}$, 
A.~Masoni$^{1}$, 
P.~Martins$^{2,5}$, 
A.~Neves$^{5}$, 
H.~Ohnishi$^{8}$, 
C.~Oppedisano$^{10}$, 
P.~Parracho$^{2}$, 
P.~Pillot$^{6}$, 
G.~Puddu$^{1}$, 
E.~Radermacher$^{2}$, 
P.~Ramalhete$^{2}$, 
P.~Rosinsky$^{2}$, 
E.~Scomparin$^{10}$, 
J.~Seixas$^{2,5}$, 
S.~Serci$^{1}$, 
R.~Shahoyan$^{2,5}$, 
P.~Sonderegger$^{5}$, 
H.J.~Specht$^{4}$, 
R.~Tieulent$^{6}$, 
G.~Usai$^{1}$, 
R.~Veenhof$^{2,5}$, 
H.K.~W\"ohri$^{2,5}$}

\runtitle{First measurement of the $\rho$ spectral function in high-energy nuclear collisions}
\runauthor{S. Damjanovic for the NA60 Collaboration}

\begin{document}

\maketitle

$^{1}$Univ.\ di Cagliari and INFN, Cagliari, Italy,
$^{2}$CERN, Geneva, Switzerland,
$^{3}$LPC, Univ.\ Blaise Pascal and CNRS-IN2P3, Clermont-Ferrand, France,
$^{4}$Univ.\ Heidelberg, Heidelberg, Germany,
$^{5}$IST-CFTP, Lisbon, Portugal,
$^{6}$IPN-Lyon, Univ.\ Claude Bernard Lyon-I and CNRS-IN2P3, Lyon, France,
$^{7}$LLR, Ecole Polytechnique and CNRS-IN2P3, Palaiseau, France,
$^{8}$RIKEN, Wako, Saitama, Japan,
$^{9}$SUNY Stony Brook, New York, USA,
$^{10}$Univ.\ di Torino and INFN, Italy,
$^{11}$YerPhI, Yerevan, Armenia

\begin{abstract}
The NA60 experiment has studied low-mass muon pairs in
158 AGeV Indium-Indium collisions at the CERN SPS. A strong excess of
pairs is observed above the expectation from neutral meson decays. The
unprecedented sample size of 360\,000 events and the good mass
resolution of about 2\% allow to isolate the excess by subtraction of
the known sources. The shape of the resulting mass spectrum is
consistent with a dominant contribution from
$\pi^{+}\pi^{-}\rightarrow\rho\rightarrow\mu^{+}\mu^{-}$
annihilation. The associated $\rho$ spectral function shows a strong
broadening, but essentially no shift in mass. 
\end{abstract}

\hspace{\fill}

Thermal dilepton production in the mass region $<$1 GeV is largely
mediated by the light vector mesons $\rho$, $\omega$ and $\phi$. Among
these, the $\rho$ is the most important, due to its strong
coupling to the $\pi\pi$ channel and its short lifetime of only 1.3
fm/c.  These properties have given it a key role as {\it the} test
particle for ``in-medium modifications'' of hadron properties close to
the QCD phase boundary. Changes both in width and in position 
were originally advocated as precursor signatures of the chiral
transition~\cite{Pisarski:mq}. There seems to be some consensus now
that the {\it width} of the $\rho$ should increase towards the
transition region, based on a number of quite different theoretical
approaches~\cite{Pisarski:mq,Dominguez:1992dw,Pisarski:1995xu,Rapp:1995zy,Rapp:1999ej}. On
the other hand, no consensus exists on how the {\it mass} of the
$\rho$ should change in approaching the transition: predictions exist
for a decrease~\cite{Pisarski:mq,Brown:2001nh}, a
constant behavior~\cite{Rapp:1995zy,Rapp:1999ej}, and even an
increase~\cite{Pisarski:1995xu}.

The CERES/NA45 experiment at the CERN SPS studied the production of
low-mass electron pairs in p-Be/Au~\cite{Agakichiev:mv},
S-Au~\cite{Agakichiev:1995xb} and Pb-Au~\cite{Agakichiev:1997au}. The
common feature of all results from nuclear collisions was an excess of
the observed dilepton yield above the known electromagnetic decays of
the produced neutral mesons, by a factor of 2-3, for masses above
0.2~GeV. The surplus yield has generally been attributed to
direct thermal radiation from the fireball, dominated by two-pion
annihilation $\pi^{+}\pi^{-}\rightarrow\rho\rightarrow l^{+}l^{-}$
with an intermediate $\rho$ which is strongly modified by the
medium. Statistical accuracy and mass resolution of the data were,
however, not sufficient to reach any sensitivity on the {\it
character} of the in-medium changes.

The new experiment NA60 at the CERN SPS has now achieved a decisive
breakthrough in this field. The apparatus is based on the muon
spectrometer previously used by NA50, and a newly added telescope of
radiation-tolerant silicon pixel detectors, embedded inside a 2.5 T
dipole magnet in the vertex region~\cite{Heuser:2003fs}. Matching of
the muon tracks before and after the hadron absorber both in {\it
coordinate and momentum} space improves the dimuon mass resolution in
the region of the light vector mesons from $\sim$80 to $\sim$20 MeV
and also decreases the combinatorial background of muons from $\pi$
and K decays. Moreover, the additional bend by the dipole field leads
to a strong increase of the detector acceptance for dimuons of low
mass and low transverse momentum. Finally, the selective dimuon
trigger and the fast readout speed of the pixel telescope allow to run
at very high luminosities.

\vspace*{-0.5cm}
\begin{figure}[htb]
\begin{minipage}[t]{80mm}
\includegraphics*[width=6cm,height=6cm,clip=,bb = 0 0 543 629]{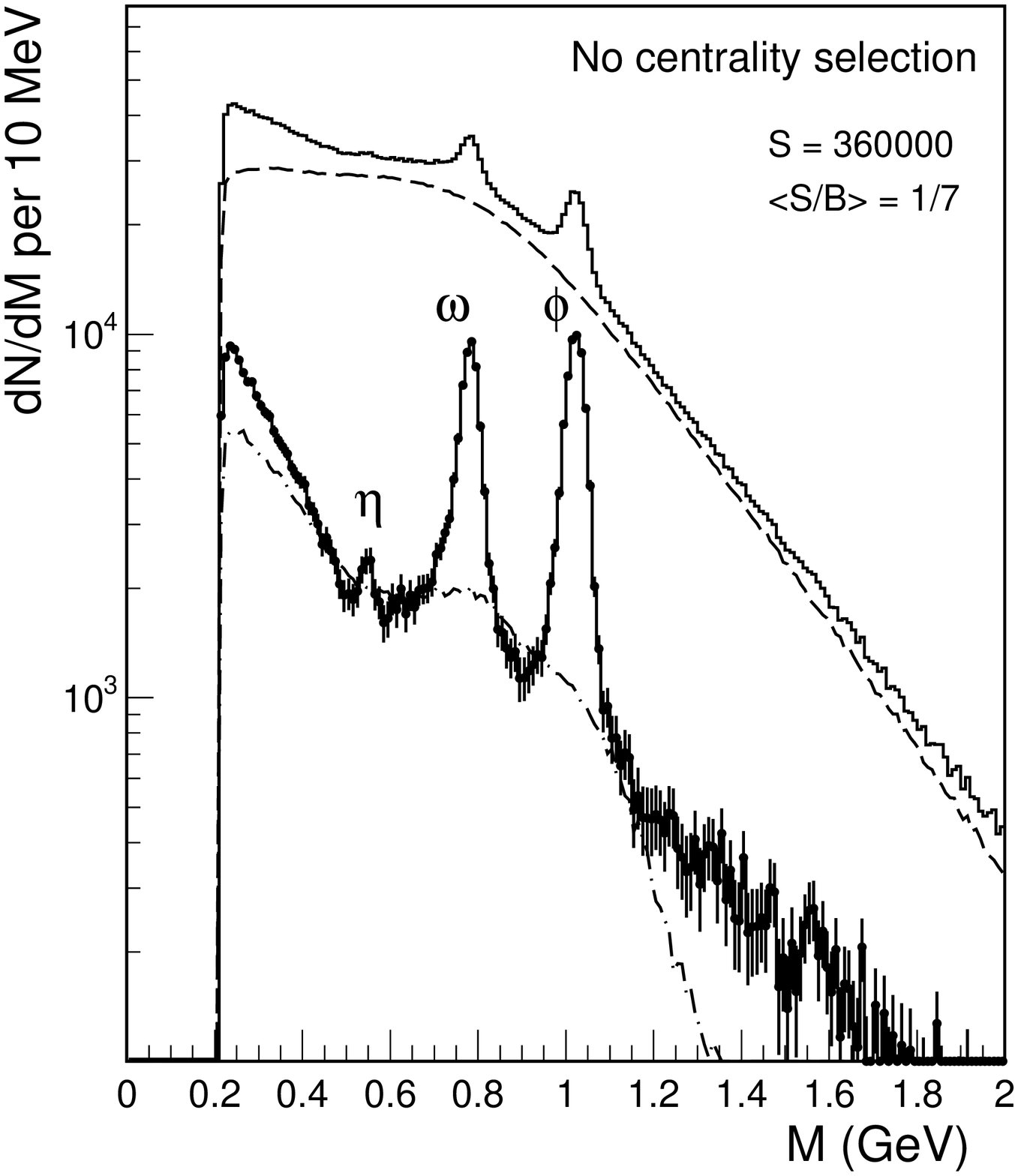}
\vspace*{-1cm}
\caption{Dimuon mass spectra of the total data (upper
histogram), combinatorial background (dashed), fake
matches (dashed-dotted), and net spectrum after subtraction of the
former two (lower)}
\label{fig1}
\end{minipage}
\hspace{\fill}
\begin{minipage}[t]{75mm}
\includegraphics*[width=6cm,height=6cm,clip=, bb= 0 0 561 649]{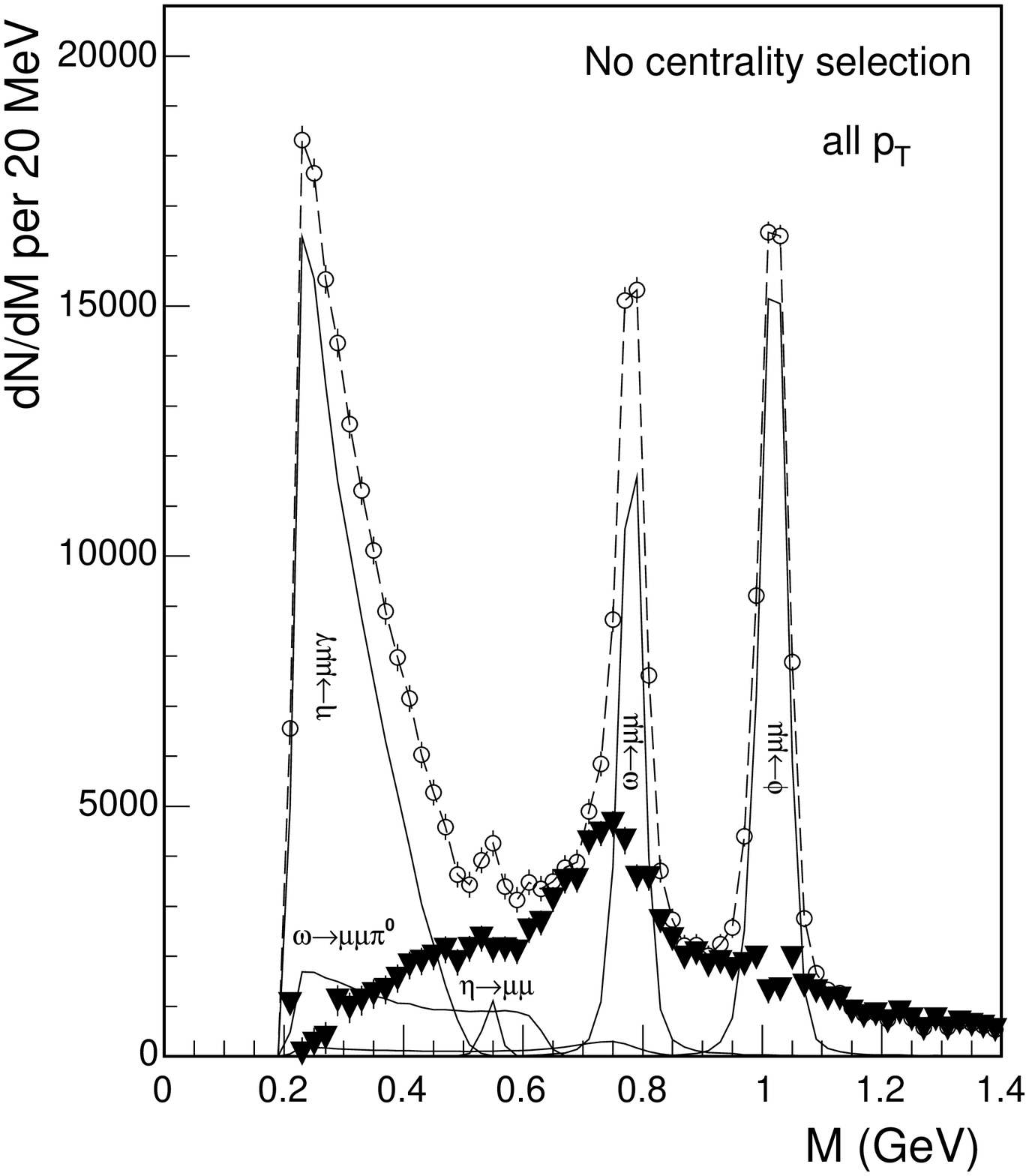}
\vspace*{-1cm}
\caption{Isolation of the excess (see text). Total data (open
circles), decay sources (solid), difference data (thick triangles),
sum of decay sources and difference data (dashed)}
\label{fig2}
\end{minipage}
\end{figure}
\vspace*{-0.5cm} 
The results reported here were obtained from the
analysis of data taken in 2003 for In-In at 158 AGeV . The main steps
of the analysis are reconstruction of the muon-spectrometer tracks,
pattern recognition and tracking in the pixel telescope including
reconstruction of the interaction vertex, and track matching. The
major background left after matching is the combinatorial background
of uncorrelated muon pairs from $\pi$ and $K$ decays. It is assessed
with a {\it mixed-event technique}~\cite{Ruben:2005qm}; the accuracy
reached is about 1\%. After subtraction of this
background, the remaining data still have to be corrected for fake
matches, i.e.~associations of muons to non-muon tracks in the
pixel-telescope. This background is minor, about 7\% of the former. It
has been estimated both with an overlay MC method and with
event-mixing~\cite{Ruben:2005qm}; the two methods agree to within 5\%.

The dimuon mass spectra associated with the different analysis steps
are shown in Fig.~\ref{fig1}. The net spectrum contains about 360\,000
pairs, with a mean signal-to-background ratio of 1/7. For the first
time in nuclear collisions, the vector mesons $\omega$ and $\phi$ are
clearly visible in the dilepton channel; even the
$\eta$$\rightarrow$$\mu$$\mu$ decay is seen. The mass resolution at
the $\omega$ is $\sim$20 MeV.  The further analysis is done in
4 different classes of collision centrality, based on the
charged-particle multiplicity distribution from the pixel
telescope. The nomenclature used below is peripheral
($dN_{ch}/d\eta$=4-30), semiperipheral~($dN_{ch}/d\eta$=30-110),
semicentral~ ($dN_{ch}/d\eta$=110-170) and
central~($dN_{ch}/d\eta$=170-240).

The peripheral data can be well described on the basis of known
sources. Muon pairs arising from the resonance ($\eta, \rho, \omega,
\phi$) and Dalitz ($\eta, \eta^{'},\omega$) decays of the produced
mesons were propagated through the NA60 set-up with GEANT, using the
generator GENESIS~\cite{Agakichiev:mv,genesis:2003} as 
input. Four free parameters were used in the fit of this ``hadron
decay cocktail'' to the data: the cross section ratios $\eta/\omega$,
$\rho/\omega$, $\phi/\omega$ and the level of charm decays.
The fits were independently done in 3 bins of dimuon
transverse momenta: p$_{T}$$<$0.5, 0.5$<$p$_{T}$$<$1 and p$_{T}$$>$1
GeV/c. The fit results for $\eta/\omega$ agree, within $<$10\%, with
the world average for pp, pBe~\cite{Agakichiev:mv}. The ratio
$\phi/\omega$ is higher by~a factor of 1.8, as to be expected due to
$\phi$ enhancement.  Both ratios are, within 10\%, independent of the
pair p$_{T}$. The results imply that the acceptance of NA60, both in
mass and p$_{T}$, is well under control.
\vspace*{-0.5cm}
\begin{figure}[h!]
\hspace{\fill}
\begin{minipage}[h!]{65mm}
\includegraphics*[width=3.2cm, height=2.5cm]{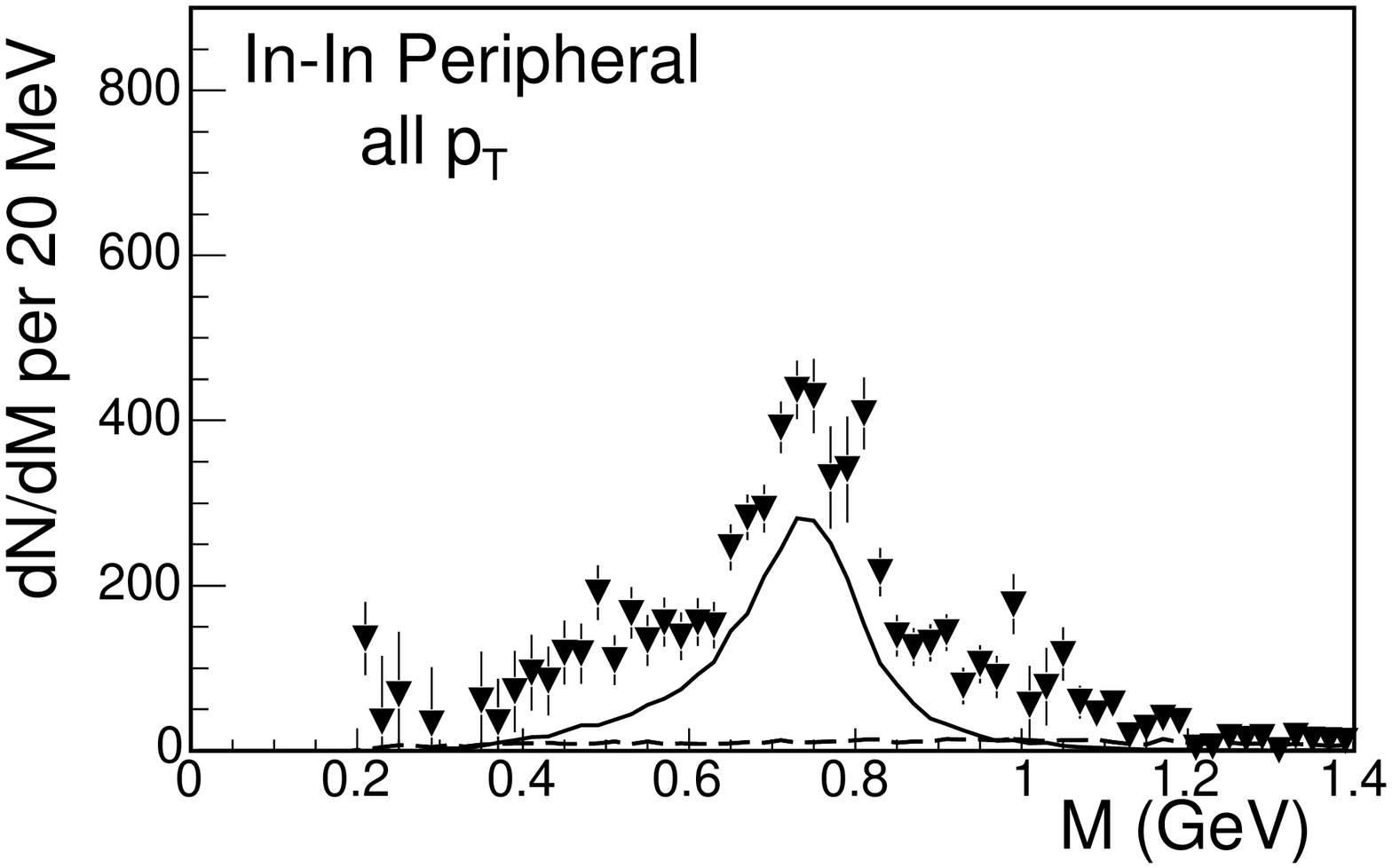}
\includegraphics*[width=3.2cm, height=2.5cm]{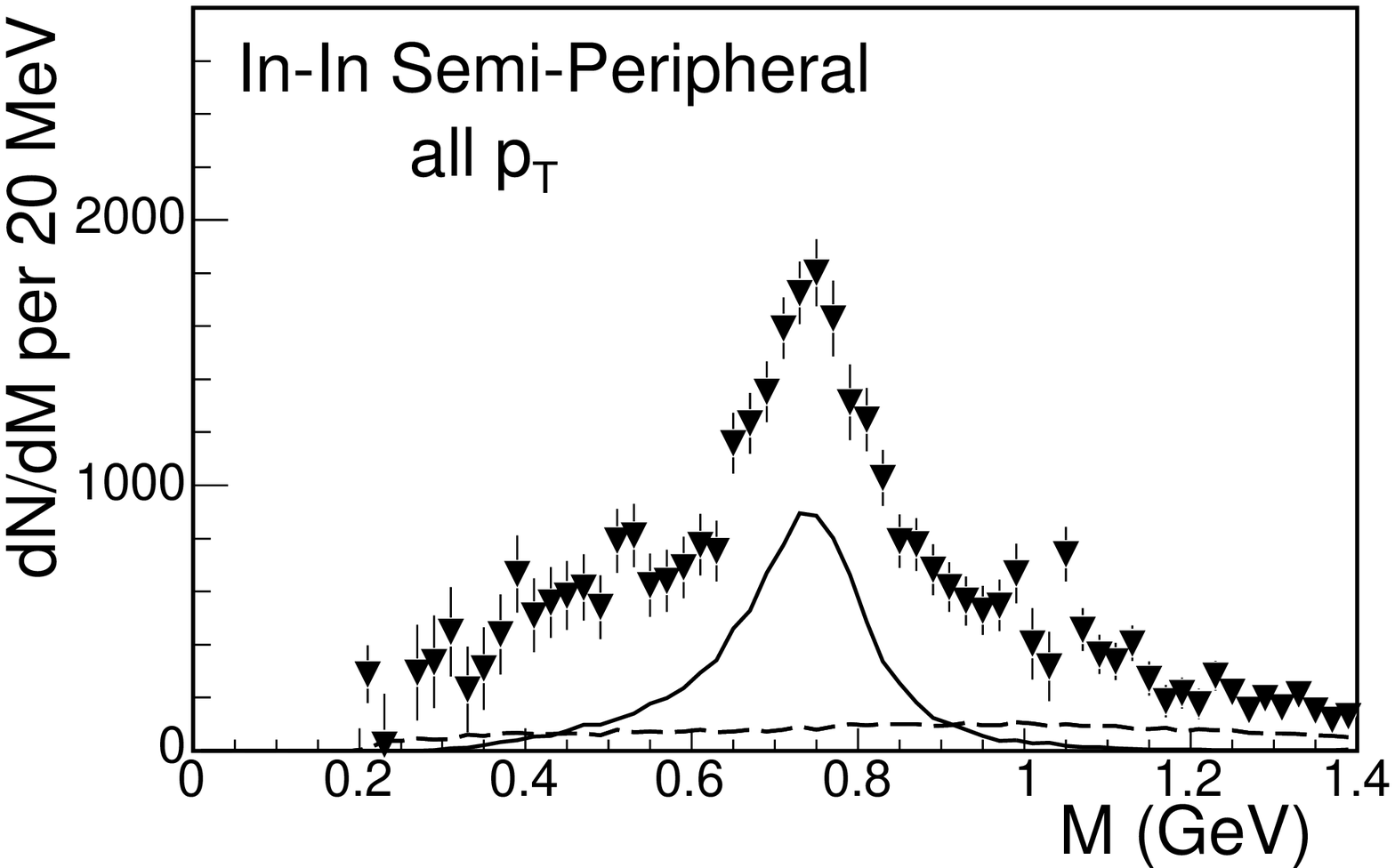}
\includegraphics*[width=3.2cm, height=2.5cm]{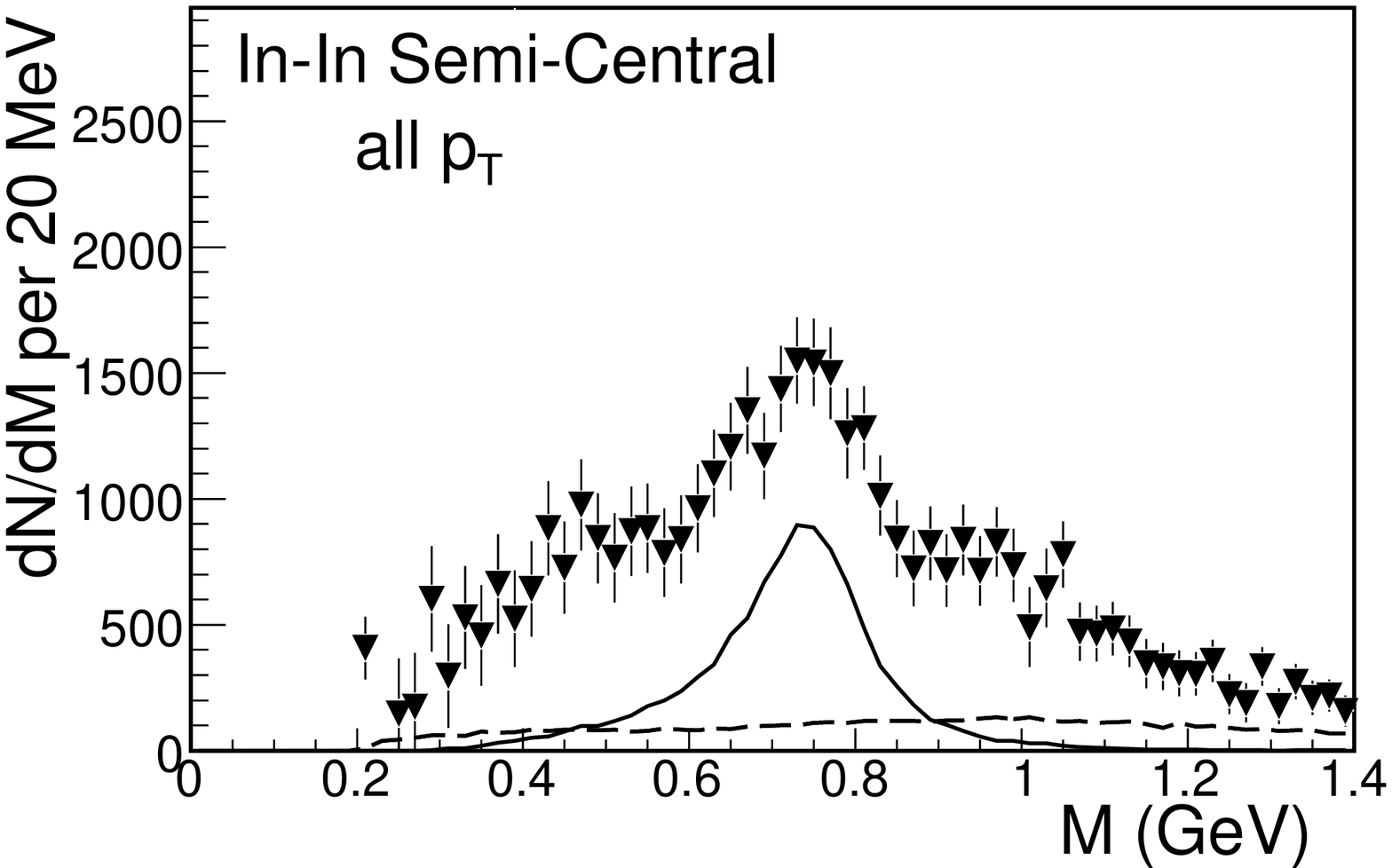}
\includegraphics*[width=3.2cm, height=2.5cm]{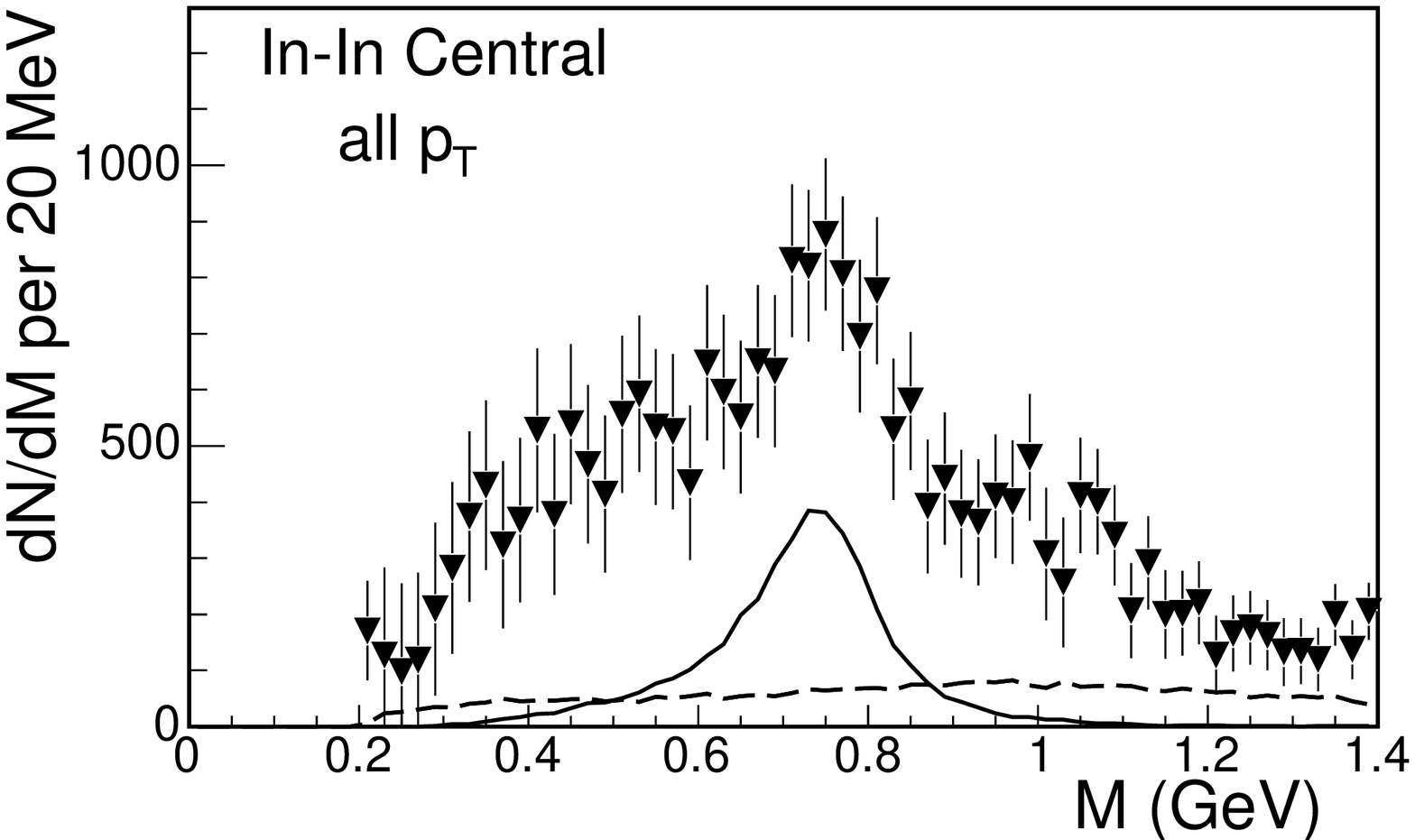}
\end{minipage}
\hspace{\fill}
\begin{minipage}[h!]{70mm}
\includegraphics*[width=6.3cm, height=5.3cm]{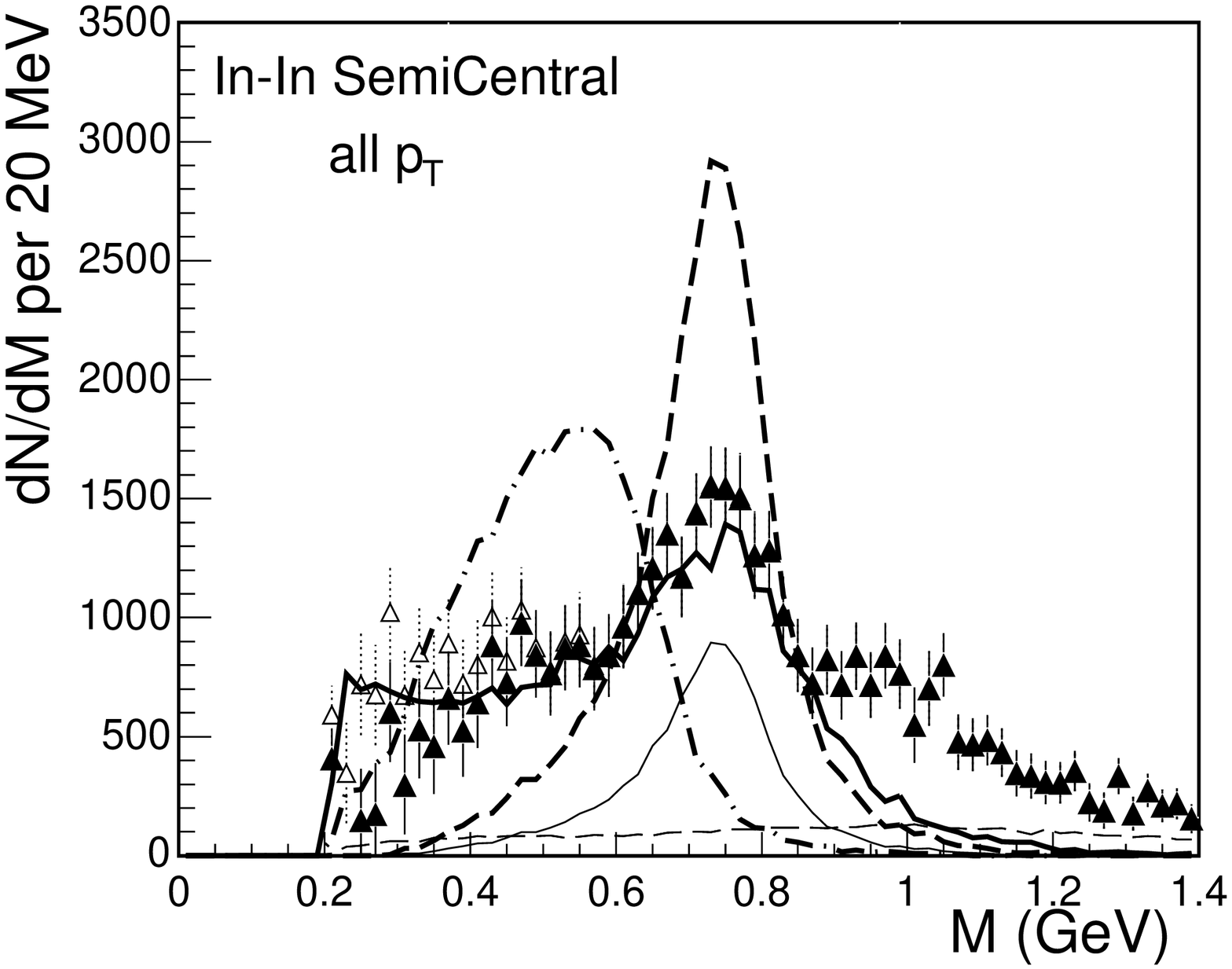}
\end{minipage}
\vspace*{-0.8cm}
\caption{Excess mass spectra of dimuons; the cocktail $\rho$ (thin
solid) and the level of combinatorial charm (thin dashed) are shown
for comparison. Left: Centrality dependence. Right: Comparison to
model predictions~\cite{rapp:nn23}, made for In-In at
$dN_{ch}/d\eta$=140. Unmodified $\rho$ (thick dashed), in-medium
broadening $\rho$ (Rapp/Wambach, thick solid), in-medium moving $\rho$
(related to Brown/Rho scaling, dashed-dotted). The open data points
reflect the change in the difference spectrum resulting from a
decrease of the $\eta$-yield by 10\%.}
   \label{fig3}
\end{figure}
\vspace*{-0.6cm}

In the more central bins, a fit procedure is ruled out, due to
the existence of a strong excess with {\it a priori unknown}
characteristics. We have, therefore, used a novel procedure as shown in
Fig.~\ref{fig2}, made possible by the excellent data quality. The
excess is {\it isolated} by subtracting the cocktail (without the
$\rho$) from the data. The cocktail is fixed, separately in each
centrality bin, by a most ``conservative'' approach. For the $\eta$,
an upper limit is defined by ``saturating'' the measured data in
the region close to 0.2 GeV; this leads to a
{\it lower} limit (zero) for the excess at very low mass, by
construction. The yields of the narrow vector mesons $\omega$ and
$\phi$ are fixed such as to get, after subtraction, a {\it smooth}
underlying continuum. The $\eta$ resonance- and $\omega$ Dalitz decays
are now bound as well.  The {\it cocktail $\rho$} (only required in
Fig.~\ref{fig3}) is bound by the ratio
$\rho/\omega$=1.2. The accuracy in the determination of the
$\eta$, $\omega$ and $\phi$ yields by this subtraction procedure is
on the level of 1-2\%, due to the remarkable {\it local} sensitivity.

The excess mass spectra for all 4 multiplicity bins are shown in
Fig.~\ref{fig3} (left). The qualitative features of the spectra are
striking: a $\rho$-like structure is seen in all cases, remaining
centered at the position of the nominal $\rho$-pole, but broadening
strongly with centrality. At the same time, the total yield increases
relative to the cocktail $\rho$, reaching values $>$4 for the most
central bin. The errors attached to the data points are purely
statistical. The systematic errors in the broad continuum region
underlying the $\rho$-like structure are estimated to be about 3\%,
12\%, 25\% and 25\% in the 4 centrality bins. The qualitative features
of these spectra are consistent with an interpretation of the excess
as dominantly due to $\pi\pi$ annihilation. In Fig.~\ref{fig3}
(right), two of the theoretical scenarios mentioned in the
introduction are confronted with the data. Note that the integrals of
the theoretical spectra are normalized to the data in the mass
interval 0.2$<$M$<$0.9 GeV. The unmodified $\rho$ and the specific
moving-mass scenario plotted are clearly ruled out. The broadening
scenario appears more realistic; for M$>$0.9 GeV, however, the data
show a nearly symmetrical broadening around the $\rho$ pole. Processes
other than 2$\pi$, i.e.\ 4$\pi$, 6$\pi$... could possibly account for
this region~\cite{rapp:nn23}.

All the data shown have not been corrected for the mass- and p$_{T}$-
dependent acceptance of the NA60 set-up. The theoretical calculations
shown in Fig.~\ref{fig3} were therefore also propagated through the
acceptance filter. It is interesting to note that, {\it by accident},
the propagation of theoretical calculations based on a white spectral
function (e.g.~$q\overline{q}$ annihilation~\cite{rapp:nn23}) yields a
white mass spectrum up to about 1 GeV. In other words, the always
existing steep rise of the theoretical input at low masses, due to the
photon propagator and a Boltzmann-like
factor~\cite{Rapp:1995zy,Rapp:1999ej,Brown:2001nh}, is just about
compensated by the falling acceptance in this region. The data and
model predictions shown in Fig.~\ref{fig3} can therefore be directly
interpreted as spectral functions of the $\rho$, averaged over momenta
and the complete space-time evolution of the fireball.  The flat part
of the measured spectra may thus reflect the early history with a nearly
divergent width, while the narrow peak on top
may just be due to the late part, approaching the nominal width.

We conclude, independent of any comparison
to theoretical modeling, that the $\rho$ primarily broadens
in In-In collisions, but does not show any noticeable shift
in mass.


\vspace*{-0.4cm}



\begin{thebibliography}{99}
\vspace*{-0.3cm}\itemsep=0cm
\bibitem{Pisarski:mq} 
R.~D.~Pisarski, Phys.\ Lett.\ {\bf 110B}, 155 (1982).
\bibitem{Dominguez:1992dw}
  C.~A.~Dominguez, M.~Loewe and J.~C.~Rojas, Z.\ Phys.\ {\bf C59}, 63 (1993).
\bibitem{Pisarski:1995xu}
  R.~D.~Pisarski, Phys.\ Rev.\ D {\bf 52}, 3773 (1995).
\bibitem{Rapp:1995zy}
R.~Rapp, G.~Chanfray and J.~Wambach, Nucl.\ Phys.\ {\bf A617}, 472 (1997).
\bibitem{Rapp:1999ej}
R.~Rapp and J.~Wambach, Adv.\ Nucl.\ Phys. {\bf 25}, 1 (2000).
\bibitem{Brown:2001nh}
G.~E.~Brown and M.~Rho, Phys.\ Rept. {\bf 363}, 85 (2002).
\bibitem{Agakichiev:mv} 
G.~Agakichiev {\em et al.} (CERES Coll.), Eur.\ Phys.\ J.\ {\bf C4}, 231 (1998).
\bibitem{Agakichiev:1995xb}
  G.~Agakichiev {\it et al.} (CERES Coll.), Phys.\ Rev.\ Lett.\  {\bf 75}, 1272 (1995).
\bibitem{Agakichiev:1997au} 
G.~Agakichiev {\it et al.} (CERES Coll.), Eur.\ Phys.\ J.\ {\bf C41}, 475 (2005).
\bibitem{Heuser:2003fs} 
G.~Usai {\it et al.} (NA60 Coll.), Eur.\ Phys.\ J.\ C {\bf 43} (2005) 415.
\bibitem{Ruben:2005qm}
R.~Shahoyan {\em et al.} (NA60 Coll.), Eur.\ Phys.\ J.\ C {\bf 43} (2005) 209, and these proceedings
\bibitem{genesis:2003} 
S.~Damjanovic, A.~De~Falco and H. W\"ohri (NA60 Coll.), Internal Note (2003).
\bibitem{rapp:nn23}
R.~Rapp, private communication (2003).
\end{thebibliography}
\end{document}